# 微服務效率分析研究
# Research on Efficiency Analysis of Microservices

**Abel C. H. Chen**

**Telecommunication Laboratories, Chunghwa Telecom Co., Ltd.**

## 摘要

隨著網路服務(Web Service)、容器(Container)、以及雲端運算(Cloud Computing)技術的成熟，傳統系統中的大服務(如：機器學習和人工智慧的計算服務)逐漸被拆分為許多個微服務來增加服務可重用性和彈性。因此，本研究提出一個效率分析框架，基於排隊模型分析傳統大服務拆分為 $n$ 個微服務的效率差異。為了讓分析結果可以泛化，本研究考慮不同的服務時間分佈(如：服務時間呈指數分佈、服務時間為固定時間等)，並且各別通過排隊模型(如：M/M/1 排隊模型和 M/D/1 排隊模型等)，探討在最差案例和最佳案例時的系統效率。在每個實驗中都顯示出原本拆分前的大服務所需花費的總時間高於拆分成多個微服務所需花費的總時間，所以拆分成多個微服務將可以提高系統效率。並且，可以觀察到在最佳案例中，隨著到達率的增加，改進效果會越顯著。然而，在最差案例中，僅有些許改進。本研究發現拆分為多個微服務後可以有效提升系統效率，並且證明當大服務的計算時間平均分配到多個微服務時，可以達到最好的改進效果。因此，本研究發現可以作為未來開發微服務架構的參考指引。

**關鍵詞：微服務、效率分析、排隊模型**

---

*通訊作者: Abel C. H. Chen (chchen.scholar@gmail.com)

## Abstract

With the maturity of web services, containers, and cloud computing technologies, large services in traditional systems (e.g. the computation services of machine learning and artificial intelligence) are gradually being broken down into many microservices to increase service reusability and flexibility. Therefore, this study proposes an efficiency analysis framework based on queuing models to analyze the efficiency difference of breaking down traditional large services into $n$ microservices. For generalization, this study considers different service time distributions (e.g. exponential distribution of service time and fixed service time) and explores the system efficiency in the worst-case and best-case scenarios through queuing models (i.e. M/M/1 queuing model and M/D/1 queuing model). In each experiment, it was shown that the total time required for the original large service was higher than that required for breaking it down into multiple microservices, so breaking it down into multiple microservices can improve system efficiency. It can also be observed that in the best-case scenario, the improvement effect becomes more significant with an increase in arrival rate. However, in the worst-case scenario, only slight improvement was achieved. This study found that breaking down into multiple microservices can effectively improve system efficiency and proved that when the computation time of the large service is evenly distributed among multiple microservices, the best improvement effect can be achieved. Therefore, this study's findings can serve as a reference guide for future development of microservice architecture.

# 壹、前言

隨著網路服務(Web Service)、容器(Container)、以及雲端運算(Cloud Computing)技術的成熟，傳統系統中的大服務逐漸被拆分為許多個微服務來增加服務可重用性和彈性(Ortiz et al., 2022; Yu et al., 2019)。因此，企業開始盤點系統程式中的耦合性和依賴關係，將可以平行處理和可重用性的程式拆分為微服務，通常 API 的方式來呼叫和處理。除此之外，也可以把需要常駐系統中執行的程式以微服務的型式在容器中常駐執行，不用常駐一個大系統耗費大量的記憶體資源和系統運算資源，可以節省系統資源。因此，如何拆分大服務成為多個微服務成為重要的議題。

為比較微服務架構和傳統架構的系統效率比較，本研究提出運用排隊模型(Samanta & Tang, 2020; Wang, Casale, & Filieri, 2023)分析傳統大服務拆分為 $n$ 個微服務的效率差異。為了讓分析結果可以泛化，本研究考慮不同的服務時間分佈(如：服務時間呈指數分佈、服務時間為固定時間等)，並且各別通過排隊模型(如：M/M/1 排隊模型和 M/D/1 排隊模型等)，探討在最差案例和最佳案例時的系統效率。通過最差案例和最佳案例的比較，可以作為拆分大服務為多個微服務的參考指引。

本論文主要分為 4 個章節。在第貳節中，描述本研究提出的效率分析框架，並且運用排隊理論和數學模型論證微服務的系統效率。在第參節中，提供數值分析，以資料圖表方式呈現效率提升效果。最後，第肆節總結本研究的貢獻和討論未來研究方向。

# 貳、研究方法

為客觀分析採用微服務架構後，對系統效率的提升程度，本研究運用排隊模型進行原理上的論證。在第貳.一節中先描述本研究提出的效率分析框架，在第貳.二節和第貳.三節分別採用 M/M/1 排隊模型和 M/D/1 排隊模型進行討論。

### 一、效率分析框架

在本研究的效率分析框架中，假設到達率呈泊松分佈(Poisson Distribution)且平均值為 $\lambda$，並且考量不同的服務時間分佈的情況下分別進行討論。為讓讀者容易閱讀，本論文以循序漸進的方式介紹，如圖 1 所示。首先，假設服務時間呈指數分佈(Exponential Distribution)，介紹當拆成 2 個微服務後的證明，以及推廣為拆成 $n$ 個微服務後的證明。然後，假設服務時間呈固定時間，介紹當拆成 2 個微服務後的證明，以及推廣為拆成 $n$ 個微服務後的證明。

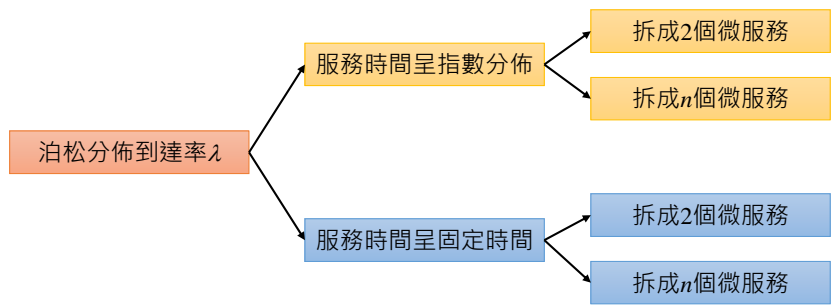

圖 1、效率分析框架

## 二、基於 M/M/1 排隊模型的效率分析

本節將假設服務時間呈指數分佈，探索拆分成 $n$ 個微服務後的效率，並且從最差案例(worst case)和最佳案例(best case)分別證明。

### (一) 拆分為 2 個微服務

**(1) 最差案例**

最差案例是指當拆分為 2 個微服務，但大部分的運算工作仍集中在其中一個微服務。因此，在本案例中假設拆分後的第 1 個微服務的平均服務時間為 $\frac{1}{\lambda+\varepsilon}$ (其中，$\varepsilon > 0$)、第 2 個微服務的平均服務時間為 $\frac{1}{\mu}$ (其中，$\mu > \lambda$)，而拆分前的大服務的平均服務時間為 $\frac{1}{\frac{\mu(\lambda+\varepsilon)}{\mu+\lambda+\varepsilon}}$ (即 $\frac{1}{\lambda+\varepsilon} + \frac{1}{\mu} = \frac{\mu+\lambda+\varepsilon}{\mu(\lambda+\varepsilon)} = \frac{1}{\frac{\mu(\lambda+\varepsilon)}{\mu+\lambda+\varepsilon}}$，並且 $\frac{\mu(\lambda+\varepsilon)}{\mu+\lambda+\varepsilon} > \lambda$)。因此，第 1 個微服務的平均服務率為 $\lambda + \varepsilon$、第 2 個微服務的平均服務率為 $\mu$、拆分前的大服務的平均服務率為 $\frac{\mu(\lambda+\varepsilon)}{\mu+\lambda+\varepsilon}$。假設上述服務時間皆服從指數分佈，可以根據 M/M/1 排隊模型得到第 1 個微服務的系統內停留時間(等待時間+服務時間)為 $T_1$、第 2 個微服務的系統內停留時間為 $T_2$、拆分前的大服務的系統內停留時間為 $T_B$，分別如公式(1)、(2)、(3)所示。

$$T_1 = \frac{1}{\lambda + \varepsilon - \lambda} = \frac{1}{\varepsilon} \tag{1}$$

$$T_2 = \frac{1}{\mu - \lambda} \tag{2}$$

$$T_B = \frac{1}{\frac{\mu(\lambda+\varepsilon)}{\mu+\lambda+\varepsilon} - \lambda} \tag{3}$$

為證明拆分為 2 個微服務後所需花費的總時間小於拆分前的大服務所需花費的總時間，則表示需證明 $T_1 + T_2 < T_B$，詳細證明如公式(4)所示。其中，根據定義 $\mu > \lambda > 0$，所以 $\lambda - \mu < 0$ 且 $\lambda - 2\mu < 0$；因此，可得 $\lambda(\lambda - \mu) + \varepsilon(\lambda - 2\mu) < 0$，得證拆分為 2 個微服務的總時間較短。

$$T_1 + T_2 < T_B$$

$$\rightarrow \frac{1}{\varepsilon} + \frac{1}{\mu - \lambda} < \frac{1}{\frac{\mu(\lambda+\varepsilon)}{\mu+\lambda+\varepsilon} - \lambda}$$

$$\rightarrow \frac{\mu - \lambda + \varepsilon}{\mu\varepsilon - \lambda\varepsilon} < \frac{\mu + \lambda + \varepsilon}{\mu\varepsilon - \lambda^2 - \lambda\varepsilon}$$

$$\rightarrow (\mu^2\varepsilon - \mu\lambda^2 - \mu\lambda\varepsilon) - (\mu\lambda\varepsilon - \lambda^3 - \lambda^2\varepsilon) + (\mu\varepsilon^2 - \lambda^2\varepsilon - \lambda\varepsilon^2) \tag{4}$$
$$< (\mu^2\varepsilon - \mu\lambda\varepsilon) + (\mu\lambda\varepsilon - \lambda^2\varepsilon) + (\mu\varepsilon^2 - \lambda\varepsilon^2)$$

$$\rightarrow \mu^2\varepsilon - \mu\lambda^2 - \mu\lambda\varepsilon - \mu\lambda\varepsilon + \lambda^3 + \lambda^2\varepsilon + \mu\varepsilon^2 - \lambda^2\varepsilon - \lambda\varepsilon^2$$
$$< \mu^2\varepsilon - \mu\lambda\varepsilon + \mu\lambda\varepsilon - \lambda^2\varepsilon + \mu\varepsilon^2 - \lambda\varepsilon^2$$

$$\rightarrow -\mu\lambda^2 - \mu\lambda\varepsilon + \lambda^3 + \lambda^2\varepsilon < \mu\lambda\varepsilon$$

$$\rightarrow \lambda^2 - \mu\lambda + \lambda\varepsilon - 2\mu\varepsilon < 0$$

$$\rightarrow \lambda(\lambda - \mu) + \varepsilon(\lambda - 2\mu) < 0, \text{where } \lambda - \mu < 0 \text{ and } \lambda - 2\mu < 0$$

**(2) 最佳案例**

最佳案例是指當拆分為 2 個微服務，拆分前的大服務的工作可以平均分配到拆分後的 2 個微服務。因此，在本案例中假設拆分後的第 1 個微服務和第 2 個微服務的平均服

務時間皆為$\frac{1}{2\mu}$，而拆分前的大服務的平均服務時間為$\frac{1}{\mu}$。因此，第 1 個微服務和第 2 個微服務的平均服務率為$2\mu$、拆分前的大服務的平均服務率為$\mu$。假設上述服務時間皆服從指數分佈，可以根據 M/M/1 排隊模型得到第 1 個微服務的系統內停留時間(等待時間+服務時間)為$T_1$、第 2 個微服務的系統內停留時間為$T_2$、拆分前的大服務的系統內停留時間為$T_B$，分別如公式(5)、(6)、(7)所示。

$$T_1 = \frac{1}{2\mu - \lambda} \tag{5}$$

$$T_2 = \frac{1}{2\mu - \lambda} \tag{6}$$

$$T_B = \frac{1}{\mu - \lambda} \tag{7}$$

為證明拆分為 2 個微服務後所需花費的總時間小於拆分前的大服務所需花費的總時間，則表示需證明$T_1 + T_2 < T_B$，詳細證明如公式(8)所示。其中，根據定義$\lambda > 0$，所以，得證拆分為 2 個微服務的總時間較短。

$$\begin{aligned}
&T_1 + T_2 < T_B \\
&\rightarrow \frac{1}{2\mu - \lambda} + \frac{1}{2\mu - \lambda} < \frac{1}{\mu - \lambda} \\
&\rightarrow \frac{2}{2\mu - \lambda} < \frac{1}{\mu - \lambda} \\
&\rightarrow 2\mu - \lambda > 2\mu - 2\lambda \\
&\rightarrow \lambda > 0
\end{aligned} \tag{8}$$

**(二) 拆分為 $n$ 個微服務**

**(1) 最差案例**

最差案例是指當拆分為$n$個微服務，但大部分的運算工作仍集中在其中一個微服務，其他$n-1$個服務各佔小部分的運算工作。因此，在本案例中假設拆分後的第 1 個微服務的平均服務時間為$\frac{1}{\lambda+\varepsilon}$ (其中，$\varepsilon > 0$)、其他第 $i$ 個微服務的平均服務時間為$\frac{1}{(n-1)\mu}$ (即$\frac{\frac{1}{\mu}}{n-1} = \frac{1}{(n-1)\mu}$，並且$\mu > \lambda$)，而拆分前的大服務的平均服務時間為$\frac{1}{\frac{\mu(\lambda+\varepsilon)}{\mu+\lambda+\varepsilon}}$ (即$\frac{1}{\lambda+\varepsilon}$ + $\sum_{i=2}^{n}\frac{1}{(n-1)\mu} = \frac{1}{\lambda+\varepsilon} + \frac{1}{\mu} = \frac{\mu+\lambda+\varepsilon}{\mu(\lambda+\varepsilon)} = \frac{1}{\frac{\mu(\lambda+\varepsilon)}{\mu+\lambda+\varepsilon}}$，並且$\frac{\mu(\lambda+\varepsilon)}{\mu+\lambda+\varepsilon} > \lambda$)。因此，第 1 個微服務的平均服務率為$\lambda + \varepsilon$、其他第 $i$ 個微服務的平均服務率為$(n-1)\mu$、拆分前的大服務的平均服務率為$\frac{\mu(\lambda+\varepsilon)}{\mu+\lambda+\varepsilon}$。假設上述服務時間皆服從指數分佈，可以根據 M/M/1 排隊模型得到第 1 個微服務的系統內停留時間(等待時間+服務時間)為$T_1$、其他第 $i$ 個微服務的系統內停留時間為$T_i$、拆分前的大服務的系統內停留時間為$T_B$，分別如公式(1)、(2)、(3)所示。

$$T_1 = \frac{1}{\lambda + \varepsilon - \lambda} = \frac{1}{\varepsilon} \tag{9}$$

$$T_i = \frac{1}{(n-1)\mu - \lambda} \tag{10}$$

$$T_B = \frac{1}{\frac{\mu(\lambda+\varepsilon)}{\mu+\lambda+\varepsilon} - \lambda} \tag{11}$$

為證明拆分為$n$個微服務後所需花費的總時間小於拆分前的大服務所需花費的總時

間,則表示需證明$\sum_{i=1}^{n} T_i < T_B$,詳細證明如公式(12)所示。其中,根據公式(4)和定義$\mu > \lambda > 0$,所以$\lambda - \mu < 0$且$\lambda - 2\mu < 0$;因此,可得$\lambda(\lambda - \mu) + \varepsilon(\lambda - 2\mu) < 0$,得證拆分為$n$個微服務的總時間較短。

$$\sum_{i=1}^{n} T_i < T_B$$

$$\to \frac{1}{\varepsilon} + \sum_{i=2}^{n} \frac{1}{(n-1)\mu - \lambda} < \frac{1}{\frac{\mu(\lambda + \varepsilon)}{\mu + \lambda + \varepsilon} - \lambda}$$

$$\to \frac{1}{\varepsilon} + \frac{n-1}{(n-1)\mu - \lambda} < \frac{1}{\frac{\mu(\lambda + \varepsilon)}{\mu + \lambda + \varepsilon} - \lambda} \quad (12)$$

$$\to \frac{1}{\varepsilon} + \frac{1}{\mu - \lambda} < \frac{1}{\frac{\mu(\lambda + \varepsilon)}{\mu + \lambda + \varepsilon} - \lambda}$$

$$\to \lambda(\lambda - \mu) + \varepsilon(\lambda - 2\mu) < 0, \text{where } \lambda - \mu < 0 \text{ and } \lambda - 2\mu < 0$$

**(2) 最佳案例**

最佳案例是指當拆分為$n$個微服務,拆分前的大服務的工作可以平均分配到拆分後的$n$個微服務。因此,在本案例中假設拆分後的第$i$個微服務的平均服務時間皆為$\frac{1}{n\mu}$,而拆分前的大服務的平均服務時間為$\frac{1}{\mu}$。因此,第$i$個微服務的平均服務率為$n\mu$、拆分前的大服務的平均服務率為$\mu$。假設上述服務時間皆服從指數分佈,可以根據 M/M/1 排隊模型得到第$i$個微服務的系統內停留時間為$T_i$、拆分前的大服務的系統內停留時間為$T_B$,分別如公式(13)、(14)所示。

$$T_i = \frac{1}{n\mu - \lambda} \quad (13)$$
$$T_B = \frac{1}{\mu - \lambda} \quad (14)$$

為證明拆分為$n$個微服務後所需花費的總時間小於拆分前的大服務所需花費的總時間,則表示需證明$\sum_{i=1}^{n} T_i < T_B$,詳細證明如公式(15)所示。其中,根據定義$\lambda > 0$,所以,得證拆分為$n$個微服務的總時間較短。

$$\sum_{i=1}^{n} T_i < T_B$$

$$\to \sum_{i=1}^{n} \frac{1}{n\mu - \lambda} < \frac{1}{\mu - \lambda} \quad (15)$$

$$\to \frac{n}{n\mu - \lambda} < \frac{1}{\mu - \lambda}$$
$$\to n\mu - \lambda > n\mu - n\lambda$$
$$\to (n-1)\lambda > 0$$

## 三、基於 M/D/1 排隊模型的效率分析

本節將假設服務時間為固定時間，探索拆分成 $n$ 個微服務後的效率，並且從最差案例和最佳案例分別證明。

### (一) 拆分為 2 個微服務

**(1) 最差案例**

最差案例是指當拆分為 2 個微服務，但大部分的運算工作仍集中在其中一個微服務。因此，在本案例中假設拆分後的第 1 個微服務的平均服務時間為 $\frac{1}{\lambda+\varepsilon}$ (其中，$\varepsilon > 0$)、第 2 個微服務的平均服務時間為 $\frac{1}{\mu}$ (其中，$\mu > \lambda$)，而拆分前的大服務的平均服務時間為 $\frac{1}{\frac{\mu(\lambda+\varepsilon)}{\mu+\lambda+\varepsilon}}$ (即 $\frac{1}{\lambda+\varepsilon} + \frac{1}{\mu} = \frac{\mu+\lambda+\varepsilon}{\mu(\lambda+\varepsilon)} = \frac{1}{\frac{\mu(\lambda+\varepsilon)}{\mu+\lambda+\varepsilon}}$，並且 $\frac{\mu(\lambda+\varepsilon)}{\mu+\lambda+\varepsilon} > \lambda$)。因此，第 1 個微服務的平均服務率為 $\lambda+\varepsilon$、第 2 個微服務的平均服務率為 $\mu$、拆分前的大服務的平均服務率為 $\frac{\mu(\lambda+\varepsilon)}{\mu+\lambda+\varepsilon}$。假設上述服務時間皆為固定時間，可以根據 M/D/1 排隊模型得到第 1 個微服務的系統內停留時間(等待時間+服務時間)為 $T_1$、第 2 個微服務的系統內停留時間為 $T_2$、拆分前的大服務的系統內停留時間為 $T_B$，分別如公式(16)、(17)、(18)所示。

$$T_1 = \frac{1}{\lambda+\varepsilon} + \frac{1}{2(\lambda+\varepsilon)}\frac{\lambda}{\varepsilon} \tag{16}$$

$$T_2 = \frac{1}{\mu} + \frac{1}{2\mu}\frac{\lambda}{\mu-\lambda} \tag{17}$$

$$T_B = \frac{1}{\frac{\mu(\lambda+\varepsilon)}{\mu+\lambda+\varepsilon}} + \frac{1}{2\frac{\mu(\lambda+\varepsilon)}{\mu+\lambda+\varepsilon}}\frac{\lambda}{\frac{\mu(\lambda+\varepsilon)}{\mu+\lambda+\varepsilon}-\lambda} \tag{18}$$

為證明拆分為 2 個微服務後所需花費的總時間小於拆分前的大服務所需花費的總時間，則表示需證明 $T_1 + T_2 < T_B$，詳細證明如公式(19)所示。其中，根據定義 $\mu > \lambda$，所以得證拆分為 2 個微服務的總時間較短。

$$\begin{aligned}
& T_1 + T_2 < T_B \\
& \to \frac{1}{\lambda+\varepsilon} + \frac{1}{2(\lambda+\varepsilon)}\frac{\lambda}{\varepsilon} + \frac{1}{\mu} + \frac{1}{2\mu}\frac{\lambda}{\mu-\lambda} < \frac{1}{\frac{\mu(\lambda+\varepsilon)}{\mu+\lambda+\varepsilon}} + \frac{1}{2\frac{\mu(\lambda+\varepsilon)}{\mu+\lambda+\varepsilon}}\frac{\lambda}{\frac{\mu(\lambda+\varepsilon)}{\mu+\lambda+\varepsilon}-\lambda} \\
& \to \frac{1}{\varepsilon(\lambda+\varepsilon)} + \frac{1}{\mu(\mu-\lambda)} < \frac{1}{\frac{\mu(\lambda+\varepsilon)}{\mu+\lambda+\varepsilon}\frac{\mu(\lambda+\varepsilon)}{\mu+\lambda+\varepsilon}-\lambda} \\
& \to \frac{\mu}{\varepsilon} + \frac{\lambda+\varepsilon}{\mu-\lambda} < \frac{(\mu+\lambda+\varepsilon)^2}{\mu\varepsilon - \lambda^2 - \lambda\varepsilon} \\
& \to \frac{\mu^2 - \mu\lambda + \lambda\varepsilon + \varepsilon^2}{\mu\varepsilon - \lambda\varepsilon} < \frac{(\mu^2+\mu\lambda+\mu\varepsilon) + (\mu\lambda+\lambda^2+\lambda\varepsilon) + (\mu\varepsilon+\lambda\varepsilon+\varepsilon^2)}{\mu\varepsilon - \lambda^2 - \lambda\varepsilon} \\
& \to (\mu^3\varepsilon - \mu^2\lambda\varepsilon + \mu\lambda\varepsilon^2 + \mu\varepsilon^3) - (\mu^2\lambda^2 - \mu\lambda^3 + \lambda^3\varepsilon + \lambda^2\varepsilon^2) \\
& \qquad - (\mu^2\lambda\varepsilon - \mu\lambda^2\varepsilon + \lambda^2\varepsilon^2 + \lambda\varepsilon^3) \\
& \qquad < (\mu\varepsilon - \lambda\varepsilon)(\mu^2 + \lambda^2 + \varepsilon^2 + 2\mu\lambda + 2\mu\varepsilon + 2\lambda\varepsilon)
\end{aligned} \tag{19}$$

$$\to \mu^3\varepsilon - \mu^2\lambda\varepsilon + \mu\lambda\varepsilon^2 + \mu\varepsilon^3 - \mu^2\lambda^2 + \mu\lambda^3 - \lambda^3\varepsilon - \lambda^2\varepsilon^2 - \mu^2\lambda\varepsilon + \mu\lambda^2\varepsilon - \lambda^2\varepsilon^2 - \lambda\varepsilon^3$$
$$< \mu^3\varepsilon + \mu\lambda^2\varepsilon + \mu\varepsilon^3 + 2\mu^2\lambda\varepsilon + 2\mu^2\varepsilon^2 + 2\mu\lambda\varepsilon^2 - \mu^2\lambda\varepsilon - \lambda^3\varepsilon - \lambda\varepsilon^3 - 2\mu\lambda^2\varepsilon - 2\mu\lambda\varepsilon^2 - 2\lambda^2\varepsilon^2$$

$$\to 3\mu\lambda\varepsilon^2 + \mu\lambda^3 + 2\mu^2\lambda\varepsilon < 3\mu^2\lambda\varepsilon + 2\mu^2\varepsilon^2 + 2\mu\varepsilon^2 + \mu^2\lambda^2$$

$$\to 3\lambda\varepsilon^2 + \lambda^3 + 2\lambda^2\varepsilon < 3\mu\lambda\varepsilon + 2\mu\varepsilon^2 + 2\lambda\varepsilon^2 + \mu\lambda^2$$

$$\to \lambda^3 < \mu\lambda^2, \text{ where } \lim_{\varepsilon\to 0}3\lambda\varepsilon^2 = 0, \lim_{\varepsilon\to 0}2\lambda^2\varepsilon = 0, \lim_{\varepsilon\to 0}3\mu\lambda\varepsilon = 0, \lim_{\varepsilon\to 0}2\mu\varepsilon^2$$
$$= 0, \lim_{\varepsilon\to 0}2\lambda\varepsilon^2 = 0$$

$$\to \lambda < \mu$$

**(2) 最佳案例**

最佳案例是指當拆分為 2 個微服務，拆分前的大服務的工作可以平均分配到拆分後的 2 個微服務。因此，在本案例中假設拆分後的第 1 個微服務和第 2 個微服務的平均服務時間皆為$\frac{1}{2\mu}$，而拆分前的大服務的平均服務時間為$\frac{1}{\mu}$。因此，第 1 個微服務和第 2 個微服務的平均服務率為$2\mu$、拆分前的大服務的平均服務率為$\mu$。假設上述服務時間皆為固定時間，可以根據 M/D/1 排隊模型得到第 1 個微服務的系統內停留時間(等待時間+服務時間)為$T_1$、第 2 個微服務的系統內停留時間為$T_2$、拆分前的大服務的系統內停留時間為$T_B$，分別如公式(20)、(21)、(22)所示。

$$T_1 = \frac{1}{2\mu} + \frac{1}{4\mu}\frac{\lambda}{2\mu - \lambda} \tag{20}$$

$$T_2 = \frac{1}{2\mu} + \frac{1}{4\mu}\frac{\lambda}{2\mu - \lambda} \tag{21}$$

$$T_B = \frac{1}{\mu} + \frac{1}{2\mu}\frac{\lambda}{\mu - \lambda} \tag{22}$$

為證明拆分為 2 個微服務後所需花費的總時間小於拆分前的大服務所需花費的總時間，則表示需證明$T_1 + T_2 < T_B$，詳細證明如公式(23)所示。其中，根據定義$\mu > 0$，所以，得證拆分為 2 個微服務的總時間較短。

$$T_1 + T_2 < T_B$$
$$\to \frac{1}{2\mu} + \frac{1}{4\mu}\frac{\lambda}{2\mu - \lambda} + \frac{1}{2\mu} + \frac{1}{4\mu}\frac{\lambda}{2\mu - \lambda} < \frac{1}{\mu} + \frac{1}{2\mu}\frac{\lambda}{\mu - \lambda}$$
$$\to \frac{1}{4\mu}\frac{\lambda}{2\mu - \lambda} + \frac{1}{4\mu}\frac{\lambda}{2\mu - \lambda} < \frac{1}{2\mu}\frac{\lambda}{\mu - \lambda} \tag{23}$$
$$\to \frac{1}{2\mu - \lambda} < \frac{1}{\mu - \lambda}$$
$$\to 2\mu - \lambda > \mu - \lambda$$
$$\to \mu > 0$$

**(二) 拆分為 $n$ 個微服務**

**(1) 最差案例**

最差案例是指當拆分為$n$個微服務，但大部分的運算工作仍集中在其中一個微服務，其他$n - 1$個服務各佔小部分的運算工作。因此，在本案例中假設拆分後的第 1 個微服

務的平均服務時間為$\frac{1}{\lambda+\varepsilon}$ (其中，$\varepsilon>0$)、其他第 $i$ 個微服務的平均服務時間為$\frac{1}{(n-1)\mu}$ (即 $\frac{\frac{1}{\mu}}{n-1}=\frac{1}{(n-1)\mu}$，並且$\mu>\lambda$)，而拆分前的大服務的平均服務時間為$\frac{1}{\frac{\mu(\lambda+\varepsilon)}{\mu+\lambda+\varepsilon}}$ (即 $\frac{1}{\lambda+\varepsilon}+\sum_{i=2}^{n}\frac{1}{(n-1)\mu}=\frac{1}{\lambda+\varepsilon}+\frac{1}{\mu}=\frac{\mu+\lambda+\varepsilon}{\mu(\lambda+\varepsilon)}=\frac{1}{\frac{\mu(\lambda+\varepsilon)}{\mu+\lambda+\varepsilon}}$，並且$\frac{\mu(\lambda+\varepsilon)}{\mu+\lambda+\varepsilon}>\lambda$)。因此，第 1 個微服務的平均服務率為$\lambda+\varepsilon$、其他第 $i$ 個微服務的平均服務率為$(n-1)\mu$、拆分前的大服務的平均服務率為$\frac{\mu(\lambda+\varepsilon)}{\mu+\lambda+\varepsilon}$。假設上述服務時間皆為固定時間，可以根據 M/D/1 排隊模型得到第 1 個微服務的系統內停留時間(等待時間+服務時間)為$T_1$、其他第 $i$ 個微服務的系統內停留時間為$T_i$、拆分前的大服務的系統內停留時間為$T_B$，分別如公式(24)、(25)、(26)所示。

$$T_1 = \frac{1}{\lambda+\varepsilon} + \frac{1}{2(\lambda+\varepsilon)}\frac{\lambda}{\lambda+\varepsilon-\lambda} = \frac{1}{\lambda+\varepsilon} + \frac{1}{2(\lambda+\varepsilon)}\frac{\lambda}{\varepsilon} \tag{24}$$

$$T_i = \frac{1}{(n-1)\mu} + \frac{1}{2(n-1)\mu}\frac{\lambda}{(n-1)\mu-\lambda} \tag{25}$$

$$T_B = \frac{1}{\frac{\mu(\lambda+\varepsilon)}{\mu+\lambda+\varepsilon}} + \frac{1}{2\frac{\mu(\lambda+\varepsilon)}{\mu+\lambda+\varepsilon}}\frac{\lambda}{\frac{\mu(\lambda+\varepsilon)}{\mu+\lambda+\varepsilon}-\lambda} \tag{26}$$

為證明拆分為$n$個微服務後所需花費的總時間小於拆分前的大服務所需花費的總時間，則表示需證明$\sum_{i=1}^{n}T_i<T_B$，詳細證明如公式(27)所示。其中，根據公式(19) 已證明結果$\frac{1}{\varepsilon(\lambda+\varepsilon)}+\frac{1}{\mu(\mu-\lambda)}<\frac{1}{\frac{\mu(\lambda+\varepsilon)}{\mu+\lambda+\varepsilon}}\frac{1}{\frac{\mu(\lambda+\varepsilon)}{\mu+\lambda+\varepsilon}-\lambda}$，而$\frac{1}{\varepsilon(\lambda+\varepsilon)}+\frac{1}{\mu((n-1)\mu-\lambda)}<\frac{1}{\varepsilon(\lambda+\varepsilon)}+\frac{1}{\mu(\mu-\lambda)}$ (詳細證明如公式(28)所示)，所以$\frac{1}{\varepsilon(\lambda+\varepsilon)}+\frac{1}{\mu((n-1)\mu-\lambda)}<\frac{1}{\frac{\mu(\lambda+\varepsilon)}{\mu+\lambda+\varepsilon}}\frac{1}{\frac{\mu(\lambda+\varepsilon)}{\mu+\lambda+\varepsilon}-\lambda}$，所以得證拆分為$n$個微服務的總時間較短。

$$\sum_{i=1}^{n}T_i < T_B$$

$$\rightarrow \frac{1}{\lambda+\varepsilon} + \frac{1}{2(\lambda+\varepsilon)}\frac{\lambda}{\varepsilon} + \sum_{i=2}^{n}\left(\frac{1}{(n-1)\mu} + \frac{1}{2(n-1)\mu}\frac{\lambda}{(n-1)\mu-\lambda}\right)$$

$$< \frac{1}{\frac{\mu(\lambda+\varepsilon)}{\mu+\lambda+\varepsilon}} + \frac{1}{2\frac{\mu(\lambda+\varepsilon)}{\mu+\lambda+\varepsilon}}\frac{\lambda}{\frac{\mu(\lambda+\varepsilon)}{\mu+\lambda+\varepsilon}-\lambda} \tag{27}$$

$$\rightarrow \frac{1}{\varepsilon(\lambda+\varepsilon)} + \frac{1}{\mu((n-1)\mu-\lambda)} < \frac{1}{\frac{\mu(\lambda+\varepsilon)}{\mu+\lambda+\varepsilon}}\frac{1}{\frac{\mu(\lambda+\varepsilon)}{\mu+\lambda+\varepsilon}-\lambda}$$

$$\frac{1}{\varepsilon(\lambda+\varepsilon)} + \frac{1}{\mu((n-1)\mu-\lambda)} < \frac{1}{\varepsilon(\lambda+\varepsilon)} + \frac{1}{\mu(\mu-\lambda)}$$

$$\rightarrow \frac{1}{\mu((n-1)\mu-\lambda)} < \frac{1}{\mu(\mu-\lambda)} \tag{28}$$

$$\rightarrow (n-1)\mu-\lambda > \mu-\lambda$$

$$\rightarrow (n-1)\mu > \mu$$

$$\rightarrow n-1 > 0$$

→ $n > 1$

**(2) 最佳案例**

最佳案例是指當拆分為 $n$ 個微服務,拆分前的大服務的工作可以平均分配到拆分後的 $n$ 個微服務。因此,在本案例中假設拆分後的第 $i$ 個微服務的平均服務時間皆為 $\frac{1}{n\mu}$,而拆分前的大服務的平均服務時間為 $\frac{1}{\mu}$。因此,第 $i$ 個微服務的平均服務率為 $n\mu$、拆分前的大服務的平均服務率為 $\mu$。假設上述服務時間皆為固定時間,可以根據 M/D/1 排隊模型得到第 $i$ 個微服務的系統內停留時間為 $T_i$、拆分前的大服務的系統內停留時間為 $T_B$,分別如公式(29)、(30)所示。

$$T_i = \frac{1}{n\mu} + \frac{1}{2n\mu}\frac{\lambda}{n\mu - \lambda} \tag{29}$$

$$T_B = \frac{1}{\mu} + \frac{1}{2\mu}\frac{\lambda}{\mu - \lambda} \tag{30}$$

為證明拆分為 $n$ 個微服務後所需花費的總時間小於拆分前的大服務所需花費的總時間,則表示需證明 $\sum_{i=1}^{n} T_i < T_B$,詳細證明如公式(31)所示。其中,根據定義 $\lambda > 0$,所以,得證拆分為 $n$ 個微服務的總時間較短。

$$\begin{aligned}
&\sum_{i=1}^{n} T_i < T_B \\
&\to \sum_{i=1}^{n} \frac{1}{n\mu} + \frac{1}{2n\mu}\frac{\lambda}{n\mu - \lambda} < \frac{1}{\mu} + \frac{1}{2\mu}\frac{\lambda}{\mu - \lambda} \\
&\to \frac{1}{\mu} + \frac{1}{2\mu}\frac{\lambda}{n\mu - \lambda} < \frac{1}{\mu} + \frac{1}{2\mu}\frac{\lambda}{\mu - \lambda} \\
&\to \frac{\lambda}{n\mu - \lambda} < \frac{\lambda}{\mu - \lambda} \\
&\to n\mu - \lambda > \mu - \lambda \\
&\to (n-1)\lambda > 0
\end{aligned} \tag{31}$$

## 參、數值分析與實驗結果

為驗證本研究提出的方法和微服務的系統效率,本節將採用數值分析的方式進行實驗討論。本研究以一個大服務拆分為 2 個微服務為例進行論證,實驗結果如圖 2 所示。其中,假設在最差案例情境下的參數值為:$\varepsilon = 2$、$\mu = 18$;在最差案例情境下的參數值為:$\mu = 2.5$。在實驗資料圖中的灰色線條表示為原本拆分前的大服務所需花費的總時間、綠色線條表示為微服務 1 所需花費的時間、藍色線條表示為微服務 2 所需花費的時間、黑色線條表示為微服務 1 和微服務 2 所需花費的總時間。在每個實驗中都顯示出灰色線條在黑色線條之上,原本拆分前的大服務所需花費的總時間高於拆分成 2 個微服務所需花費的總時間,所以拆分成 2 個微服務將可以提高系統效率。

其中,可以觀察到在最佳案例(如圖 2(b)和圖 2(d))中,隨著到達率的增加,改進效果會越顯著。然而,在最差案例(如圖 2(a)和圖 2(c))中,僅有些許改進。有鑑於此,未

來在拆分大服務成為多個微服務，拆分的微服務若能平均分配原本的計算量，將可以顯著提升系統效率。

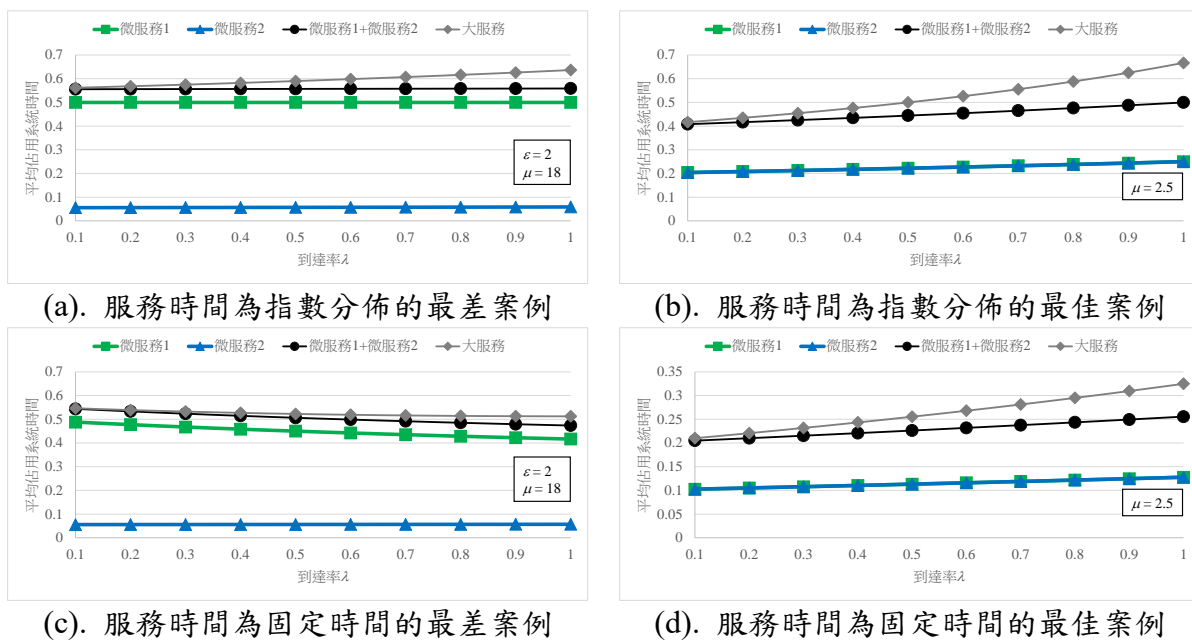

(a). 服務時間為指數分佈的最差案例　　(b). 服務時間為指數分佈的最佳案例

(c). 服務時間為固定時間的最差案例　　(d). 服務時間為固定時間的最佳案例

圖 2、大服務拆分為 2 個微服務的數值分析比較圖

## 肆、結論與未來研究建議

本研究提出一個效率分析框架，通過假設服務時間基於不同的假設前提下進行推導，讓研究成果在解釋意義上可以具備泛化性。本研究討論了把一個大服務拆分為多個微服務的系統效率，並且通過數學模型和實驗結果證明拆分成微服務可以提升系統效率。除此之外，本研究討論最差案例和最佳案例，可以證明當大服務的計算時間平均分配到多個微服務時，可以達到最好的改進效果，作為未來開發微服務架構的參考指引。

## 參考文獻